\documentclass[conference]{IEEEtran}

\usepackage{engord}
\usepackage{url,amsmath,amssymb,cite}
\usepackage{subfigure}
\usepackage{color}
\usepackage{amsmath}
\usepackage{amsmath}
\usepackage{graphicx}
\usepackage{graphics}
\usepackage[table]{xcolor} 
\usepackage{multirow}
\usepackage{rotating}
\usepackage{url}
\usepackage{amssymb}
\usepackage{algorithm}
\usepackage[noend]{algpseudocode} 

\definecolor{BgGray}{gray}{0.9}%
\definecolor{BgGray2}{gray}{0.96}%
\definecolor{RowColorOdd}{named}{BgGray2}%
\definecolor{RowColorEven}{named}{white}%
\definecolor{comments}{gray}{.5}

\newcommand\colheadbegin{\hline\rowcolor{BgGray}}
\newcommand\colheadend{\\\hline}

\begin{document}

\title{On Frequency-Selective Scheduling in SDMA-OFDMA Systems}

\author{\IEEEauthorblockN{Anatolij Zubow and Johannes Marotzke}
\IEEEauthorblockA{Humboldt University Berlin\\
\{zubow,marotzke\}@informatik.hu-berlin.de}
}

\maketitle

\begin{abstract}
Orthogonal Frequency Division Multiple Access (OFDMA) is a multi-user version of the Orthogonal Frequency Division Multiplexing (OFDM) transmission technique, which divides a wideband channel into a number of orthogonal narrowband subchannels, called subcarriers. An OFDMA system takes advantage of both \emph{frequency diversity} (FD) gain and \emph{frequency-selective scheduling} (FSS) gain. A FD gain is achieved by allocating a user the subcarriers distributed over the entire frequency band whereas a FSS gain is achieved by allocating a user adjacent subcarriers located within a subband of a small bandwidth having the most favorable channel conditions among other subbands in the entire frequency band. 
Multi-User Multiple Input Multiple Output (MU-MIMO) is a promising technology to increase spectral efficiency. A well-known MU-MIMO mode is Space-Division Multiple Access (SDMA) which can be used in the downlink direction to allow a group of spatially separable users to share the same time/frequency resources.

In this paper, we study the gain from FSS in SDMA-OFDMA systems using the example of WiMAX. Therefore, a complete SDMA-OFDMA MAC scheduling solution supporting both FD and FSS is proposed. The proposed solution is analyzed in a typical urban macro-cell scenario by means of system-level packet-based simulations, with detailed MAC and physical layer abstractions. By explicitly simulating the MAC layer overhead (MAP) which is required to signal every packed data burst in the OFDMA frame we can present the overall performance to be expected at the MAC layer.

Our results show that in general the gain from FSS when applying SDMA is low. However, under specific conditions, small number of BS antennas or large channel bandwidth, a significant gain can be achieved from FSS.
\end{abstract}


\section{Introduction}\label{sec-introduction}

Future wireless networks are faced with an ever increasing demand for higher data rates and better quality of service. Orthogonal Frequency Division Multiple Access (OFDMA) and Multi-User Multiple Input Multiple Output (MU-MIMO) are promising technologies to increase current spectral efficiencies in 4G wireless communication systems. 

In OFDMA, the entire bandwidth is divided into multiple orthogonal subcarriers, of which a subset of subcarriers is assigned to a user. In a typical cellular network the total bandwidth is usually much larger than the coherence bandwidth of the channel resulting in frequency-selective channel gains of orthogonal subcarriers.
An OFDMA system can benefit from frequency-selectivity in two different ways~\cite{Lee2007}. The first approach, called \emph{frequency diversity} (\emph{FD}), is achieved by allocating a user the subcarriers widely distributed over the entire frequency band. The performance gain from \emph{FD} is obtained by using multiple subcarriers whose path gains are independently faded rather than using subcarriers with similar faded path gains. The second approach, called \emph{frequency-selective scheduling} (\emph{FSS}), is achieved by allocating a user adjacent subcarriers located within a subband of a limited bandwidth having the most favorable channel conditions among other subbands in the entire frequency band. Note, that to achieve \emph{FSS} gain channel state information at the transmitter is required.
In WiMAX IEEE 802.16-2009~\cite{802.16-2009} both approaches are supported in the form of subchannelization, i.e. adjacent or distributed subcarrier allocation.

A well-known MU-MIMO mode is Space-Division Multiple Access (SDMA) which can be used in the downlink direction (DL) to allow a group of spatially separable MSs to share the same time/frequency resources. Users to be grouped in a SDMA group must be carefully selected since any inter-user interference will reduce the performance significantly. A SDMA-OFDMA system has to allocate resources in time, frequency and space dimensions to different MSs resulting in a highly complex resource allocation problem.

The main contribution of this paper is the evaluation of the gain from \emph{FSS} in SDMA-OFDMA based systems using the example of WiMAX. Therefore, a complete SDMA-OFDMA MAC scheduling solution for WiMAX systems supporting both \emph{FD} and \emph{FSS} is proposed.  
Therefore, we focus on both the \emph{SDMA group formation} and the \emph{OFDMA frame construction} problem. The \emph{SDMA group formation} problem considers the creation of spatially compatible SDMA groups in a frequency-selective channel, where each group consists of a set of stations which can be served by the BS using the same time and frequency domain. Due to frequency selectivity the channel transfer matrix is different for different subcarriers; hence the spatial compatibility among MSs in an SDMA group depends on the used subcarriers. From the theoretical point of view the optimal strategy is to perform SDMA grouping per subcarrier, i.e. narrow-band~\cite{Maciel2008a}. However, this is not feasible in practice, because every packed data burst in an OFDMA frame need to be signaled by an entry in the MAP resulting in considerable MAC layer overhead\footnote{A MAP is a broadcast which is required to be correctly received by all the MSs scheduled in a frame. Thus the most robust MCS must be used for transmission.}. Consequently, the amount of signaling needed to inform MSs about resources allocated to them might be substantially reduced when allocating larger chunks of time/frequency blocks for MSs.
The \emph{OFDMA frame construction} problem considers the creation of the actual OFDMA frame to be transmitted, i.e. given a set of SDMA groups we need to select a subset of SDMA groups to be scheduled in the OFDMA frame and allocate for each group a certain space within the frame. In addition, we also need to allocate a certain amount of space for signaling MAP messages used to reference the packed data bursts in the frame.

The proposed solution is analyzed in a typical urban macro-cell scenario by means of simulations. By explicitly simulating the MAC layer overhead (MAP) which is required to reference each packed data burst we can present the overall performance to be expected at the MAC layer.

The rest of this paper is structured as follows. Sections~\ref{sec:relatedwork} summarize the most relevant work in the area. Next, Section~\ref{sec-problem} formalizes the problem description and formulates the SDMA-OFDMA scheduling as an optimization problem. Section~\ref{sec-description} describes our proposed solution along with the designed algorithms required. In Section~\ref{sec-perf_eval} a performance evaluation is conducted through comprehensive simulations. Finally, Section~\ref{sec-conclusions} summarizes the main results and concludes the paper.

\section{Related Work}\label{sec:relatedwork}

To the best of the authors' knowledge all proposals in the state of the art evaluate the MAC performance of SDMA in OFDMA-based systems using \emph{FD} only. However, no analysis is available examining the gain from \emph{FSS} in a SDMA-OFDMA system.

Nascimento et al.~\cite{Nas10ajoint} proposed a joint utility packet scheduler and SDMA-based resource allocation architecture for 802.16e. They suggest a TDD frame for SDMA implementation using DL-PUSC mode. DL-PUSC is a distributed subcarrier allocation scheme hence achieving a \emph{FD} gain. The frame was sub-divided into a set of 15 radio allocation units, each comprised of 10 subchannels and 6 OFDM symbols which result in a total of 30 slots available per resource. The SDMA zone comprises the first row of resources, i.e. the first 10 subchannels. The remaining two rows are allocated for non-SDMA transmissions. To capture frequency selective channels the authors proposed the Exponential Effective SIR Mapping (EESM~\footnote{\url{www.ieee802.org/16/tge/contrib/C80216e-05_141r3.pdf}}) compression method.

In our prior work~\cite{Zubow2012a} we proposed a complete SDFMA-OFDMA scheduling solution for 802.16e. Our allocation scheme forces an allocated SDMA group to span always an integer number of columns in the OFDMA frame, i.e. an SDMA groups utilizes all subcarriers of a given time slot and hence achieves a \emph{FD} gain. This design decision allows the SDMA grouper to be agnostic to the actual frequency region where the data will be transmitted.
Furthermore, this design decision turns the packing of SDMA groups within an OFDMA frame into a simpler one-dimensional packing problem.

\section{Problem Statement}\label{sec-problem}

This section describes the physical and QoS models used throughout this paper and formulates our proposed SDMA-OFDMA scheduling solution as an optimization problem. 

\begin{figure}
   \begin{center}
       \includegraphics[width=0.6\linewidth]{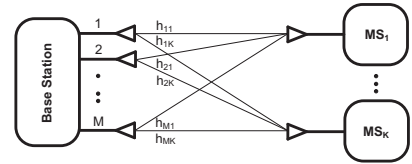}
   \end{center}
   \caption{System model: DL of a single BS equipped with Antenna Array and multiple single-antenna MSs.}
   \label{fig:system_model}
\end{figure}

\subsection{PHY Model}\label{sec-systemmodel}

We consider the downlink of a single BS with a maximum transmit power that is being equally distributed among the served MSs. The BS is equipped with an Uniform Linear Array (ULA) and there are $K$ single-antenna MSs associated with the BS (Fig.~\ref{fig:system_model}). We consider OFDMA where the channel bandwidth is divided into $S$ orthogonal subcarriers. In addition, physical subcarriers are mapped to logical subchannels using PUSC-ASCA~\cite{802.16-2009}, i.e. adjacent subcarrier allocation.
Regarding CSI knowledge in the BS, we assume that for each MS the BS knows the channel transfer matrix ($H_{u,f}$) of every $D$th subcarrier, hence resulting in $\frac{S}{D}$ $H_{u,f}$ coefficients per MS, where $S$ is the number of subcarriers. In addition, we assume that the BS has full CSI on each $h_{u,f}$, i.e. there are no estimation errors. The required channel coefficients are generated with the WIM simulator~\cite{WIMSIM}, where only low to medium-speed mobility is considered.

Next, we introduce the format used for the channel transfer matrix, the beamforming weights, and the SINR calculation.
For an MS $u$ ($u=1, \ldots ,K$) the channel coefficient $h_{i,u}$ denotes the sampled frequency response of the channel between the BS antenna element $i$ and the receiver antenna of the $u$-th MS. Thus, the channel coefficients belonging to a given MS $u$ can be grouped into a column vector $M \times 1$, i.e. $H_u = [h_{1,u},h_{2,u},...,h_{M,u}]^{\mathrm{T}}$, where $H_u$ holds all spatial correlations and multi-path effects of the MS $u$ channel. The BS uses precoding weights to adjust the transmitted signal in order to mitigate the propagation effects of the channel and to control the interference among MSs in a SDMA group. Therefore, every channel coefficient $h_{i,j}$ is associated with a beamforming weight $w_{i,j}$, and $W_u = [w_{1,u},w_{2,u},...,w_{M,u}]^{\mathrm{T}}$ is the normalized weight vector for a given MS $u$.

The effective channel gain at the BS can then be computed as $||W_u^{\mathrm{H}} \cdot H_u||_2^2$, where $(\cdot)^{\mathrm{H}}$ denotes the conjugate transpose of a vector/matrix. Notice, that multiple MSs being served during the same time-frequency resource cause interference upon each other (intra-cell interference). Therefore, given a certain frequency resource $b=1, \ldots , B$, and the corresponding channel coefficients, the SINR of MS $u$ can be computed as~\cite{Maciel2008a}:
\begin{align}\label{eq:sinr}
\gamma_{u, b} &= \frac{P_{u, b} \cdot ||W_{u, b}^{\mathrm{H}} \cdot H_{u,b}||_2^2}{\sigma^2 + \sum_{v=1, u \neq v}^M P_{u, b} \cdot ||W_{v,b}^{\mathrm{H}} \cdot H_{u,b}||_2^2}
\end{align}
where $P_{u, b}$ represents the average received power at MS $u$ on frequency resource $b$. We assume that the BS uses the \emph{Minimum Mean Square Error} technique (MinMSE ~\cite{Gross2005}) in order to compute beamforming weights.

%
%
\subsection{Base Station Architecture and QoS Model}\label{subsec:bsarch}

We consider an \emph{offline} BS architecture as formally defined in \cite{Zubow2010}. In an offline architecture there are two main building blocks that cooperate in order to maximize the utility of the scarce radio resources, these blocks are the \emph{QoS Scheduler} and the \emph{SDMA-OFDMA} scheduler. The task of the QoS Scheduler is to select from the higher layer packets available in each flow's buffer a \emph{candidate} list of packets to be transmitted in the next DL subframe. In addition, the QoS scheduler tags each individual packet $P_{i}$ with a \emph{utility} value $u_{i}$. Typical QoS schedulers that can be accomodated in this architecture are Proportional Fair or Deficit Round Robin \cite{Lakkakorpi2008}. Thus, after receiving a set of candidate packets from the QoS Scheduler, the task of the SDMA-OFDMA scheduler is to design the SDMA-OFDMA frame layout, i.e. assign the time, frequency and space resources, in order to maximize the amount of utility carried in the SDMA-OFDMA frame.

\subsection{Problem Formulation}\label{sec-optimization}

From the theoretical point of view the optimal SDMA-OFDMA resource allocation strategy in a frequency-selective channel is to perform SDMA grouping per subcarrier, i.e. narrow-band~\cite{Maciel2008a}. However, from a practical point of view this is not reasonable. Every packed data burst needs to be signaled by an entry in the DL-MAP resulting in substantial management overhead. Thus we have a tradeoff between exploiting gain from \emph{FSS} and efficiency in terms of MAP-overhead.

The main objective of a DL SDMA-OFDMA scheduler is to maximize the total utility carried in the DL subframe. However, optimally assigning time, frequency and space resources is a very complex task to be efficiently implemented in practice. Therefore, in order to reduce complexity we divide the overall optimization problem into two independent major tasks:
\begin{itemize}
 \item [1.] \emph{OFMDA frame partitioning}: we consider the problem of dividing the OFDMA frame into multiple subbands each consisting of contiguous/adjacent subcarriers.
 \item [2.] \emph{SDMA group formation}: we consider the problem of creating spatially compatible SDMA groups for each subband, where each group consists of a set of stations which can be separated in the space domain. 
 \item [3.] \emph{OFDMA frame construction}: given a set of SDMA groups per subband, we consider the problem of scheduling the subset of these groups within the OFDMA frame that maximizes the overall utility carried in the OFDMA frame.
\end{itemize}


\subsubsection*{OFMDA frame partitioning}

\begin{figure}
   \begin{center}
       \includegraphics[width=0.8\linewidth]{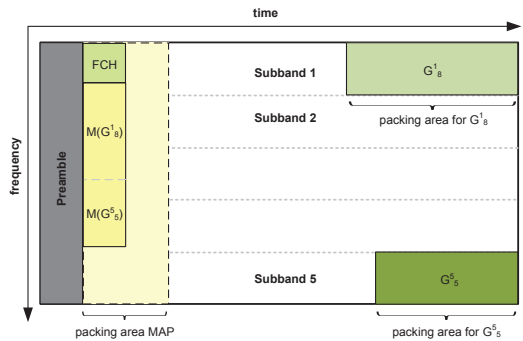}
   \end{center}
   \caption{Proposed partitioning of the OFDMA DL frame into subbands.}
   \label{fig:sdma_fss_framelayout}
\end{figure}

In a wideband or frequency selective channel the characteristics of the spatial channel cannot longer be described by a single channel transfer matrix. Due to frequency selectivity the channel transfer matrix is different for different subcarriers; hence the spatial compatibility among MSs in an SDMA group depends on the used subcarriers. In order to address this fact, the OFDMA DL frame is subdivided into multiple subbands each consisting of a particular number of adjacent subcarriers with a similar channel transfer function (Fig.~\ref{fig:sdma_fss_framelayout}).


\subsubsection*{SDMA group formation}

For each subband the SDMA grouping is performed separately. Possible frequency selectivity within a subband is captured by using the Exponential Effective SIR Mapping (EESM) compression method that allow to collapse a set of per-subcarrier SINR measurements into an effective SINR measurement, that can then be used to calculate a grouping metric like group capacity. The objective of this problem can be formalized as presented in~\cite{Zubow2012a} with the exception that the optimal SDMA grouping must be found for each subband resulting in a set of SDMA groups per subband.

\subsubsection*{OFDMA frame construction}

The task of the OFDMA frame construction is to schedule a subset of SDMA groups within the OFDMA frame so that the overall utility carried in the OFDMA frame is maximized. Due to the SDMA grouping per subband it is possible that the same MS is scheduled in different subbands sharing a SDMA group together with different MSs. It must therefore be guaranteed that a given packet is packed in at most one subband to avoid duplicate packets.

The following design decision regarding the frame construction was made. We enforce an allocated SDMA group to span always an integer number of columns of the corresponding subband. The main reason for this decision is that it enables the previous SDMA group formation problem to be agnostic to the actual frequency region where the data will be transmitted, i.e. the required SINR values can be derived assuming that an allocation spans the whole frequency range in the corresponding subband. In addition, this design decision turns the packing of SDMA groups within a subband into a one-dimensional packing problem instead of a two-dimensional packing problem. 

The OFDMA frame construction problem can be formalized as presented in~\cite{Zubow2012a} with the exception that the optimal packing area for each SDMA group in each subband need to be find.

\section{Proposed Solution}\label{sec-description}

In this section we present our proposed \emph{Frequency-Selective SDMA-OFDMA MAC Scheduler} (\emph{FOSSOS}). In order to implement the assignment of time, frequency and space resources in an efficient way, the proposed solution addresses sequentially the following three sub-problems: i) first, the problem of dividing the frame into multiple subbands to achieve a possible FSS gain, ii) second, the problem of creating spatially compatible SDMA groups in each subband, and iii) third the problem of scheduling a given set of SDMA groups within an OFDMA frame.

\begin{figure}
   \begin{center}
       \includegraphics[width=0.8\linewidth]{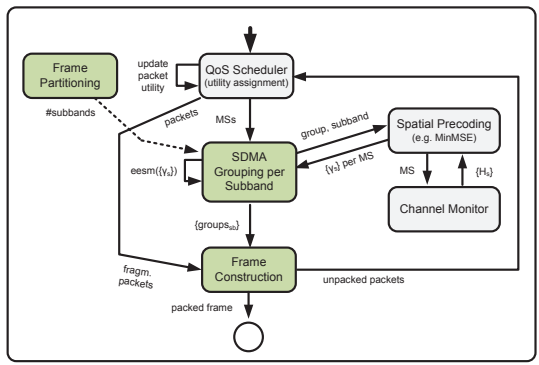}
   \end{center}
   \caption{Overview of the proposed scheduling solution.}
   \label{fig:sGSAoverview}
\end{figure}

Fig.~\ref{fig:sGSAoverview} gives an overview. In order to address the problem of creating spatially compatible SDMA groups, our solution subdivides the OFDMA DL frame into multiple subbands each consisting of a certain number of adjacent subcarriers. Partial Usage of Subchannels - Adjacent Subcarrier Allocation (PUSC-ASCA) is used. 
For each such subband the SDMA grouping is performed separately. The information about the per-subband SDMA grouping is used by the \emph{Frame construction module} to schedule the best SDMA groups in each subband and hence exploiting the gain from FSS. 


\subsection{Frame Partitioning}

The task of this module is to divide the OFDMA frame into multiple subbands (i.e. frequency resource blocks). 
The number of subbands to be used is a configurable parameter. By using a large number of subbands a higher gain from FSS can be achieved. The downside is the increased DL-MAP overhead because each packed SDMA group (2D burst) need to be referenced in the DL-MAP. This can become especially expensive when packing a great number of large SDMA groups. Thus there is a tradeoff between gain achieved by FSS and efficiency in terms of DL-MAP overhead.

\subsection{SDMA Grouping}

The SDMA grouping is performed for each subband separately. A state of the art SDMA grouper like BFA, FFA or our proposed CBA~\cite{Zubow2012a} can be used. Possible frequency selectivity within a subband is captured by using the EESM compression method.

\subsection{Frame Construction}

The proposed low-complexity algorithm constructs an OFDMA frame from multiple per-subband available SDMA groups and their packets with the main goal to maximize the total utility carried by an OFDMA frame. Additionally it supports a variable number of subbands to be used with the goal to achieve a FSS gain. The higher the total number of subbands available, the higher the probability to find a subband with favorable channel conditions for an SDMA group.

Given a set of SDMA groups per subband, the algorithm selects a subset of these groups to be allocated within the current SDMA-OFDMA frame depending on the utility of corresponding packets. In addition the space needed for signaling the DL-MAP, which varies depending on the selected groups, is allocated as well. 

Algorithm~\ref{alg:framePackingAlg} describes the algorithm consisting of two phases: the \emph{extension} phase (outer loop) and the \emph{selection} phase (inner loop). The \emph{extension} phase allocates the frame in the time dimension in a stepwise fashion, whereas the selection phase chooses the best groups overall subbands as to achieve a FSS gain.

In the \emph{extension} phase a certain area is made available for packing. In the \emph{selection} phase the best SDMA group over all subbands, which maximizes the sum utility of the frame using the available packing area is scheduled with its given size. For each next iteration within the \emph{selection} phase the best group of all \emph{remaining} subbands is scheduled. After all subbands have been allocated the \emph{extension} phase increases the total packing area in width and the \emph{selection} process is repeated, either increasing the area of an already scheduled group or adding a new one to the corresponding subband. The algorithm terminates in case there is no more space left within the frame for the \emph{extension} phase to increase the packing area for any given SDMA group.

It should be noted that the SDMA groups of different subbands are obviously overlapping, which means that the same packets can be allocated in different groups. The \emph{frame} structure used in Algorithm \ref{alg:framePackingAlg} was used for simplification and should provide the necessary functionality as to freeze and unfreeze packets depending on whether or not they have been already scheduled on another subband. In other words, once an SDMA group is added to the frame (line 18) the selected packets that fit into the current space must be prohibited from being allocated on another subband (group). Only the group which has allocated packets in a previous iteration is allowed to reuse, or unfreeze them for others. Furthermore, the amount of packets which can be packed in the packing area of a SDMA group is determined by the space available in the MAP.

\begin{algorithm*}
\footnotesize
\caption{\small The SDMA-OFDMA packing algorithm partitions a frame into different areas each representing the packing area of given SDMA group.}
\label{alg:framePackingAlg}
\begin{algorithmic}[1]
\Require \\
$\mathcal{G}$ - set of SDMA groups over all subbands; 
$P$ - list of packets destined to MSs in $G$ sorted according to utility per slot;\\
$\mathrm{ofdmaFrame}()$ - Creates a frame object which handles the frame layout and any packing related features.\\
$\mathrm{SCSB}$ - Number of subchannels per subband.\\
$\mathrm{SB}$ - Total number of subbands.
\Procedure{$\mathrm{framePart}$}{$G$, $P$}
\State $\mathrm{stepSize} \leftarrow \lceil\frac{\mathrm{minSlotSize}(G,P)}{\mathrm{SCSB}}\rceil \times \mathrm{SCSB}$
\State $\mathrm{vLimit} \leftarrow \mathrm{max}(\mathrm{initSz} \times \mathrm{SCSB}, \mathrm{stepSize})$
\State $\mathrm{usedSpace} \leftarrow [0 \cdots 0_{\mathrm{SB}}]$ \Comment{Packed space per subband}
\State $\mathrm{frame} \leftarrow \mathrm{ofdmaFrame}(P)$ \Comment{Initialize structure representing the frame}
\While{$\mathit{vLimit} \times \mathrm{SB} + \mathrm{frame}.\mathrm{MapSizeSlots}() < \mathrm{frame}.\mathrm{Size}$}
   \For{$G_i^j \in G$} \Comment{For each group on every subband}
    \State $G_i^j \leftarrow \mathrm{vLimit} - \mathrm{usedSpace}(j) + \mathrm{frame}.\mathrm{getSize}(G_i^j)$ \Comment{Set the available size}
  \EndFor  
  \State $J \leftarrow \{1 \cdots \mathrm{SB} \} $;
  \While{$|J| \neq 0}$
    \State $G_i^j \leftarrow \arg\max_{G_i^j} \mathrm{util}(\mathrm{frame} \cup G_i^j), j \in J, G_i^j \in G$
    \If {$\mathrm{util}(\mathrm{frame} \cup G_i^j) > \mathrm{util}(\mathrm{frame})$} \Comment{Ensure gain in terms of total utility}
      \State $\mathrm{frame}.\mathrm{set}(G_i^j)$ \Comment{Add or update the allocated space for the found group}
      \State $J \leftarrow J \setminus j$ \Comment{Remove all groups of subband $j$}
      \State $\mathrm{usedSpace}(j) \leftarrow \mathrm{usedSpace}(j) + \mathrm{stepSize}$
    \Else
      \State $ J \leftarrow \emptyset$ \Comment{Stop searching if we cannot increase the total utility}
    \EndIf
  \EndWhile 
  \State $\mathrm{stepSize} \leftarrow \mathrm{min}(\mathrm{max}(\mathrm{frame}.\mathrm{freeCols}, 1) \times \mathrm{SCSB}, \mathrm{stepSize})$
  \State $\mathrm{vLimit} \leftarrow \mathrm{vLimit} + \mathrm{stepSize} $ \Comment{Adjust the vertical limit}
\EndWhile 
\\
\Return ($\mathrm{frame}$)
\EndProcedure \Statex
\end{algorithmic}
\end{algorithm*}

\subsubsection*{Example}

\begin{figure}
   \begin{center}
       \includegraphics[width=1\linewidth]{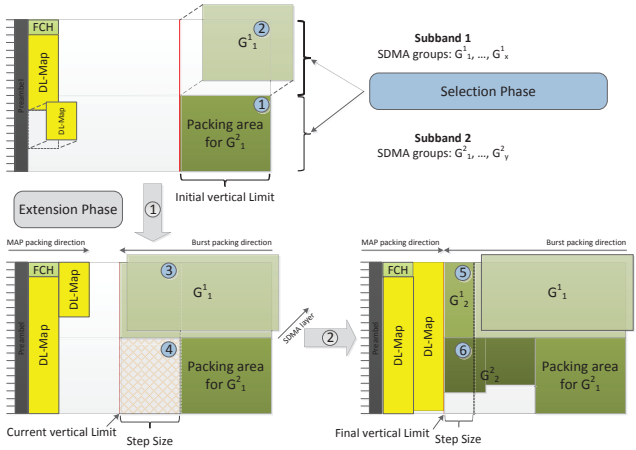}
   \end{center}
   \caption{SDMA-OFDMA packing example. The selection phase schedules SDMA groups per subband based on the maximum increase of the total utility carried by the frame. In case (4) the additional space is not sufficient to pack additional packets, stays therefore unallocated and can be scheduled during the next extension phase (6). The packing area gradually approaches the MAP area until it becomes a limiting factor for the last added SDMA groups.}
   \label{fig:subframePackingEx}
\end{figure}

Fig.~\ref{fig:subframePackingEx} illustrates a detailed example of the packing algorithm. As can be seen the extension phase moves a vertical limit for the burst packing area from right to left towards a column wise growing MAP growing from left to right. The initial vertical limit sets the largest available space for the selection phase, after which it is incremented by a predetermined \emph{step size} that eventually reduces to one single column until MAP and packed area cover all columns of the whole frame. In the last step the last scheduled SDMA groups are limited by the lack of available space for the MAP (ref. to group $G_2^2$ in Fig.~\ref{fig:subframePackingEx}).

\subsubsection*{Some more Details}

The vertical limit was a necessary design choice due to the column wise growing MAP. It ensures that the packing area of all subbands grow more evenly in the end so that no single subband blocks the MAP to early due to favorable channel conditions. The initial vertical limit as well as the step wise incrementation are an essential part of that process. The \emph{initial} vertical limit should be as large as possible in favor of the \emph{FSS} gain, as well as to minimize the eventually needed steps that are necessary for a gradual approach between the MAP and the packed area. The \emph{initial} vertical limit in full columns is calculated as follows:
\begin{equation}
 \mathit{initSz} = \Big \lceil \frac{(\mathit{DL}_{\mathit{sl}}-1) \times \mathit{SC} /\mathit{MSB} - \mathit{Map}_{\mathit{sz}}^* \times M}{\mathit{SC}} \times \mathit{SB} \Big \rceil,
\end{equation}
where $M$ is the number of antennas at BS, $\mathit{DL}_{\mathit{sl}}$ the total number of slots (i.e. columns) and $\mathit{SC}$ the total number of subchannels (i.e. rows) in the OFDMA frame. $Map_{sz}^*$ is the predicted MAP size in slots per SDMA layer and $\mathit{SB}$ the number of subbands used. $\mathit{Map}_{\mathit{sz}}^*$ was estimated based on the total number of 40 Byte packets that can be merged and signaled into $SC$ many slots on an average MCS. The idea behind this concept is that the initial vertical limit allows the same number of slots for any number of subband with $\mathit{MSB}$ being the maximum and that the pessimistically estimated large MAP allows the rest of the frame to be used for an SDMA group. This value is set in Algorithm \ref{alg:framePackingAlg} line 7. The step size should be as small as possible incrementing the packing area by one column per step. In case the packet queue is filled with very large packets a small step size might be too small for additional packets, therefore, it is overridden by the maximum minimum burst size encountered in all MS packet queues rounded to full columns (line 6). As the incrementation of the packet area approaches the map area the step size will be limited by the remaining available space (line 25).

\subsubsection*{Complexity}
There are $K$ MSs per subband each corresponding to a spatial layer, independent of SDMA group affiliation. This means the expression in line 16 has to estimate the utility gain of $K \times \mathit{SB}$ many spatial layers. The selection phase is performed $\mathit{SB}$ many times and the extension phase is at most $\mathit{DL}_{\mathit{sl}}$. The worst case total number of performed utility estimations is, therefore, $\mathcal{O}(\mathit{DL}_{\mathit{sl}} \times \mathit{SB} \times ( K \times  \mathit{SB}))  = \mathcal{O}(\mathit{SB}^2)$, i.e. quadratic with the number of subbands.

%
%
\section{Performance Evaluation}\label{sec-perf_eval}

The performance of the proposed solution is analyzed in a multi-cell environment by means of simulations according to the methodology described in~\cite{Koc2008}. The most important parameters are summarized in Table~\ref{table:simparams}. The following traffic model is used. The BS maintains per-MS packet buffers, and packets are generated based on a distribution derived from a data collection campaign by SPRINT\footnote{https://research.sprintlabs.com/packstat/}. The obtained packet size distribution is dominated by TCP flows. We do not considered the offered load to be equally distributed among all active MSs, but instead consider a more unbalanced situation where 50\% of the MSs generate 80\% of the traffic at the MAC layer. Next, the Proportional Fair Utility (PF) was used to ensure fairness among MSs. Finally, the DL-MAP advertised to the MSs is fully simulated to ensure the correct amount of signaling overhead.

\begin{table}[t]
\footnotesize
\centering
\begin{tabular}{p{.4\linewidth}p{.5\linewidth}}
\colheadbegin \textbf{Parameter} & \textbf{Value} \colheadend
System bandwidth& 10\,MHz\\
Subcarrier bandwidth& 10.9375\,kHz\\
FFT size& 1024\\
Center frequency& 2.5\,GHz\\
Frequency reuse pattern& 3x1x1\\
Transmit power& 46\,dBm\\
MS noise density (dBm/Hz)& -167 dBm/Hz\\
Cell radius& $\sim 288\,m$\\
WIM scenario& C2 (urban macro-cell, LOS/NLOS)\\
Number of antennas at BS ($M$)&  1-5 omni elements separated by half wavelength\\
No. of single-antenna MSs ($K$)& 12, 24, 36\\
MSs placement& uniform\\
MSs' speed& 2\,km/h\\
OFDMA frame duration& 5\,ms\\
DL/UL ratio& 35/12\\
MAP& 1 repetition, QPSK~1/2\\
Tx Precoding algorithm& MinMSE\\
Deployment (reuse pattern)& 3x1x1\\
WiMAX permutation scheme & PUSC-ASCA\\
Packet buffer size& $K \times 12.96$\,KiB\\
No. of placement seeds& 256\\
\end{tabular}
\caption{Simulation Parameters.}
\label{table:simparams}
\end{table}

%
%
\subsection{Simulation Results}\label{subsec:simresults}
In the following we analyze the impact of channel bandwidth, number of users, number of antennas at BS and channel propagation conditions (LOS/NLOS) on the performance of the proposed solution. For each scenario we show results when using different numbers of subbands. The aim is to show that the optimal number of subbands to be used depends on the scenario. 

\subsubsection{Impact of Bandwidth}

In Fig.~\ref{fig:thr_bw} we analyzed the impact from the used channel bandwidth, i.e. 5, 10 and 20\,MHz. Furthermore, the number of antennas at the BS, $M$, was varied from 2, 4 and 8, whereas the total number of MSs, $K$, was fixed to 12 and the LOS propagation was used.
The following observations can be made. We cannot gain from FSS when using a small channel bandwidth (5\,MHz). By increasing the number of subbands to be used for FSS the throughput decreases mainly due to increased signaling MAP overhead 
(Fig.~\ref{fig:omap_bw}). The loss is especially severe when using more antennas, e.g. with $M=8$ antennas and 6 subbands the throughput decreases by one-fifth compared to the baseline with only a single subband where \emph{FD} is achieved. This is mainly due to a two-third increase in MAP size. The gain from FSS is at most when using a small number of antennas together with a large channel bandwidth, e.g. we can achieve a 16.8\% throughput gain compared to baseline when using FSS with 6 subbands with 2 antennas and a bandwidth of 10\,MHz. Finally, the gain from FSS is higher when using a larger bandwidth.


\subsubsection{Impact of User Number}

Next we varied the total number of MSs, $K$, from 12, 24 to 36. The channel bandwidth was fixed to 10\,MHz and LOS propagation was used. From Fig.~\ref{fig:thr_usr} we can see that by increasing the number of MSs the gain from FSS increases slightly. E.g. with $M=8$ antennas the throughput gain from FSS can be increased from only 1.3\% to 5.4\% when having a network with 36 instead of 12 only MSs.

\subsubsection{Impact of LOS}
\begin{figure*}[ht]
  \begin{center}
   \mbox
    {
      \subfigure[~\emph{Impact of Bandwidth on Throughput}]
      {
    \label{fig:thr_bw}
        \scalebox{1}
        {
            \includegraphics[width=0.45\linewidth]{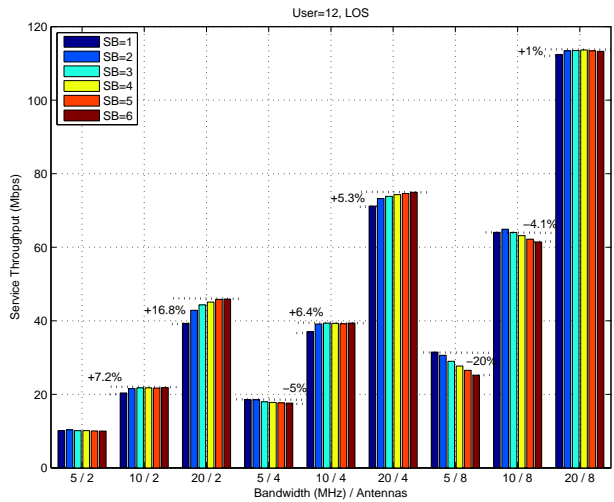}
        }
      }
      \subfigure[~\emph{Impact of Bandwidth on DL-MAP Overhead}]
      {
    \label{fig:omap_bw}
        \scalebox{1}
        {
            \includegraphics[width=0.45\linewidth]{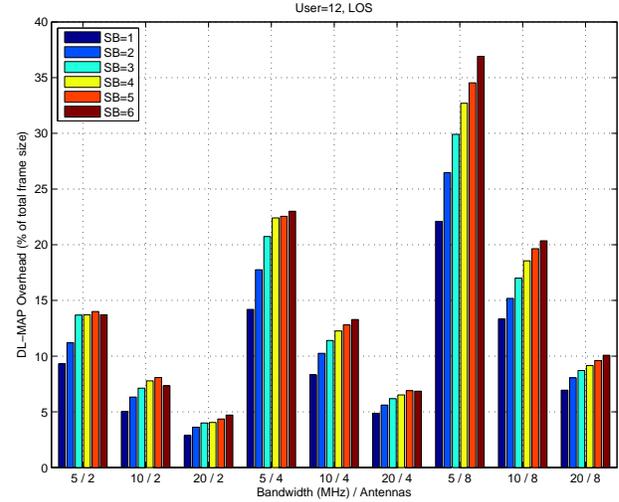}
        }
      }
    }
   \mbox
    {
      \subfigure[~\emph{Impact of User Number}]
      {
    \label{fig:thr_usr}
        \scalebox{1}
        {
            \includegraphics[width=0.45\linewidth]{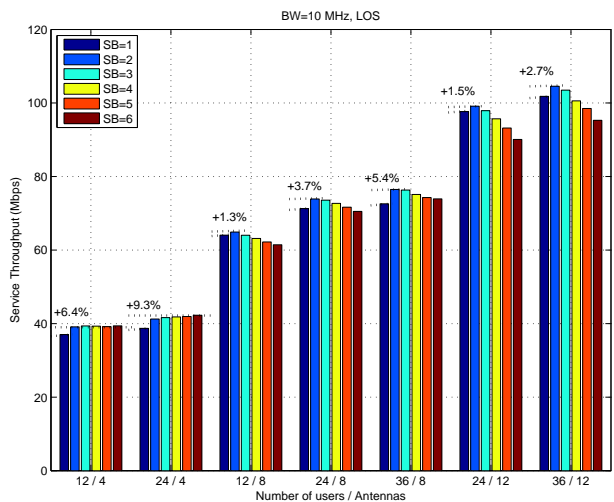}
        }
      }
      \subfigure[~\emph{Impact of LOS}]
      {
    \label{fig:thr_los}
        \scalebox{1}
        {
            \includegraphics[width=0.45\linewidth]{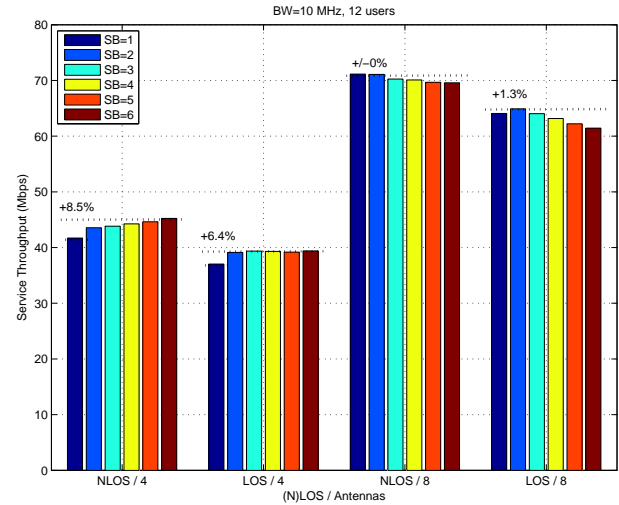}
        }
      }                      
    }
\caption{Simulations results.}
    \label{fig:perf_all}
  \end{center}
\end{figure*}

Finally, we are analyzing whether the gain from FSS is higher in a NLOS environment. From Fig.~\ref{fig:thr_los} we can see that for 4 antennas at the BS the gain from FSS is a little bit higher for NLOS (8.5\% vs. 6.4\%). With 8 antennas there is only a small gain from FSS for the LOS scenario which is again due to the high MAP overhead.

\subsubsection{Summary}

In summary, we found the following rule of thumb for the evaluated urban macro-cell environment (WIM scenario C2). There is no gain from FSS when using a channel bandwidth of 5\,MHz or smaller. For a channel bandwidth of 10\,MHz a gain from FSS can be achieved when using a small number of antennas, i.e. 2 or 4. With 20\,MHz of bandwidth there is also a FSS gain when using 8 antennas at BS.

\section{Conclusions and Future Work}\label{sec-conclusions}

In this paper we evaluated the gain from \emph{frequency-selective scheduling} (\emph{FSS}) in a SDMA-OFDMA based system using the example of WiMAX. By explicitly simulating the MAC layer overhead (i.e. DL-MAP), which is required to signal every packed data burst in the OFDMA frame, we can present the overall performance to be expected at the MAC layer. The evaluation was performed by means of system-level simulations. Our results for a typical urban macro-cell environment show that in general the gain from FSS when applying SDMA is low. However, under specific conditions, small number of BS antennas or large channel bandwidth, a significant gain of up to 16.8\% can be practically achieved from FSS.

\bibliographystyle{IEEEtran}
\bibliography{IEEEabrv,literature}

\end{document}